\documentclass[preprint,3p,twocolumn,pra,aps,showpacs,superscriptaddress,article]{elsarticle}
\usepackage[english]{babel}
\usepackage[utf8]{inputenc}
\usepackage{amsmath,amssymb,amsthm}
\usepackage{xcolor}
\usepackage{float}
\usepackage{textcomp}
\usepackage{amssymb}
\usepackage{graphicx}
\usepackage{esint}
\usepackage{lineno,hyperref}
\usepackage[toc,page]{appendix}
\usepackage{tabularx}
\def\<{\left\langle}
\def\>{\right\rangle}
\def\({\left(}
\def\){\right)}
\def\dis{\textrm{dis}}
 \journal{Physica A}

\begin{document}

\begin{frontmatter}

\title{Avalanches in an extended Schelling model: an
  explanation of urban gentrification}

\author{Diego Ortega\corref{cor1}}
\ead{dortega144@alumno.uned.es}
\address{Dto. Física Fundamental, Universidad Nacional de Educación
a Distancia (UNED), Spain}

\author{Javier Rodríguez-Laguna}
\address{Dto. Física Fundamental, Universidad Nacional de Educación
a Distancia (UNED), Spain}

\author{Elka Korutcheva}
\address{Dto. Física Fundamental, Universidad Nacional de Educación
  a Distancia (UNED), Spain}
\address{G. Nadjakov Institute of Solid State Physics, Bulgarian
  Academy of Sciences, 1784 Sofia, Bulgaria.}

\begin{abstract}
In this work we characterize sudden increases in the land price of certain urban areas, a phenomenon causing gentrification, via an extended Schelling model. An initial price rise forces some of the disadvantaged inhabitants out of the area, creating vacancies
which other groups find economically attractive. Intolerance issues
forces further displacements, possibly giving rise to an avalanche. We
consider how gradual changes in the economic environment affect the
urban architecture through such avalanche processes, when agents may
enter or leave the city freely. The avalanches are characterized by
power-law histograms, as it is usually the case in self-organized
critical phenomena.
\end{abstract}

\end{frontmatter}


\section{Introduction}

{\em Gentrification} \cite{key-10p} is an social phenomenon that takes
place when local land prices suffer a sudden increase. Real state
investors tend to expand its area in their search for economical
profit, thus creating pressure over the less economically favoured
classes, who are forced to move out and favoring segregation. A major
contribution in this field was made by T.C. Schelling \cite{key-1p},
who considered two different social groups, which we may call {\em
  red} and {\em blue}, located on a square lattice with some vacancies
on it. Agents are characterized by a certain {\em tolerance} $T$: the
fraction of different agents in their neighborhood that he or she can
tolerate. For any agent, if the fraction of diverse neighbors is lower
or equal than $T$, the agent is {\em happy} and remains at his/her
location. Otherwise, he/she relocates to the nearest vacancy that
meets his/her demands. For intermediate values of $T$ a segregated
state is reached.

Despite its simplicity, based in the idea that alike people tend to
group, the model presents complex dynamics, thus drawing a substantial
attention both from social and statistical mechanics researchers,
sometimes with different emphasis. In the latter category we find
analogues of thermodynamic quantities such as specific heat or
susceptibility, as calculated in \cite{key-6p}, the static and dynamic
properties of the Schelling model in one and two-dimensional systems
\cite{key-3p} and the interfacial roughening between clusters and
diffusion mechanisms in \cite{key-4p}. From the social perspective we
find works where agents are able to consider their future happiness
perspectives by taking into account the neighborhood evolution
\cite{key-25p}, or the influence of altruistic behaviour, given that
some agents are able to decrease their {\em happiness} for the greater
good \cite{key-26p}, having a great impact on the final state reached
by the system. The balance between cooperative and individual dynamics
was analyzed in \cite{key-27p}.

Other works characterize different features of the model, such as the
effect of the city shape, size and form \cite{key-5p}. Three regimes
(segregated, integrated and mixed) were found when the Schelling model
is extended to include agents who can considerate different
neighborhood sizes \cite{Laurie}. Furthermore, an open city model in
which agents can leave or enter the system was described in
\cite{key-7p}. In addition to showing different kinds of interfaces
between clusters, economic aspects of the system were introduced by
means of a chemical potential. Recently, the use of different
tolerance levels for the agents was proposed in \cite{key-8p}, in a
system with no vacancies, where agents could only exchange locations
with agents of a different type. On the other hand, in \cite{key-9p}
each cell of the system is considered to be a building containing many
agents, and segregation was considered both at a microscopic and a
macroscopic level, giving rise to a complex phase diagram.

In this article we consider an open city model in which the system
becomes gradually more hostile towards one type of agent and more
attractive towards the other, giving rise to a partial or total
overtake of the favored type of agent, which may proceed through
avalanches, as it has been reported in previous works. These
avalanches present a power-law behavior which is usual of similar
processes \cite{perbak,bartolozzi}. Our general framework is
established in similarity with the Blume-Emery-Griffiths (BEG) model
in presence of an external magnetic field \cite{key-6p,key-7p}.

The paper is organized as follows. In Section \ref{sec:model} we
define our BEG model and discuss the dynamics and the evolution
process. In Section \ref{sec:results} we describe our results,
highlighting the association between the overtaking of one kind of
agents over the other and the gentrification phenomena. Our main
conclusions and proposals for further work are discussed in section
\ref{sec:conclusion}.


\section{Model}
\label{sec:model}

The Blume-Emery-Griffith model \cite{key-11p} was introduced to study
the behaviour of He$^3$-He$^4$ mixtures. In this model the spin
values considered are $s_i=0,\pm 1$. In the presence of a magnetic
external field, the Hamiltonian can be written as:

\begin{equation}
  \mathcal{H}=-\sum_{\<i,j\>}\(J\,s_{i}s_{j}
  +K\,s_{i}^{2}s_{j}^{2}\)
  +\sum_i \(D_{B}\,s_{i}^{2}+H_{B}\,s_{i}\),
\label{eq:1}
\end{equation}
where $\<i,j\>$ stands for the eight nearest neighbors of a Moore
neighborhood. This Hamiltonian represents a spin-1 Ising model with
coupling constant $J$, biquadratic exchange constant $K$, crystal
field $D_B$ and an external magnetic field of intensity $H_B$. The
crystal field $D_B$ acts as a chemical potential that controls the
entry of cells with non-zero spin value. Dissimilar entry fluxes
for $s=+1$ and $s=-1$ are obtained by means of the magnetic field,
$H_B$. 

In our interpretation of the BEG model spin values will be associated
with {\em blue} agents ($s_i=+1$), {\em red} agents ($s_i=-1$) and
vacancies ($s_i=0$). A positive value of $J$ yields a negative energy
for each pair of neighboring agents of the same type, while a positive
value of $K$ assigns a negative energy to every pair of neighboring
agents, disregarding their type. If $D_B>0$, the system reduces its
energy by expelling agents, and if $H_B>0$ , the system reduces its
energy either expelling {\em blue} agents or attracting {\em red}
ones. Under certain conditions, the Hamiltonian provided by
Eq. \eqref{eq:1} can only decrease along the actual dynamical
trajectories of the system, thus serving as a Lyapunov function
\cite{key-12p}.

In order to make an explicit connection between our physical model and
social realities, let us now define a measure of the level of
unhappiness of an agent in relation with the parameters of
Eq. \eqref{eq:1}. The lack of happiness of agent $i$ is measured by
the {\em dissatisfaction} index $I_\dis(i)$

\begin{equation}
  I_\dis(i)=N_d(i)-T[N_s(i)+N_d(i)]+ D + H(i),
  \label{eq:2}
\end{equation}
where $N_s$ and $N_d$ are, respectively, the number of neighboring
agents of the same (s) and different (d) type, $D$ is a measure of the
global economic level and is the same for all agents. Meanwhile
$H(i)=+H$ for blue agents and $-H$ for red ones. Red agents can be
considered as a group of people with promising economical
perspectives. Meanwhile, the situation is opposite for blue
ones. Thus, $H$ can be understood as the economic bias, or half the
economic difference gap between both types of agents. We will drop the
dependency on $i$ when it is clear from context. The condition for
satisfaction will be, therefore,

\begin{equation}
  I_\dis(i)\leq 0.
  \label{eq:2bis}
\end{equation}

Notice that when $D<0$ the system is generally friendly towards
agents, and unhappiness arising from their neighborhood can be
endured. This is the common situation behind immigration processes:
increased economic opportunities compensate the lack of
homogeneity. On the other hand, when $D>0$, the economic opportunities
have disappeared and the system becomes hostile. Even when $N_s>N_d$,
an agent might be forced to leave the system under such
circumstances. Moreover, the possibility that economic perspectives of
both communities are not similar is taken into account by the
parameter $H$. When its absolute value is large, agents of one type
might be forced to leave due to the absence of economic stimulus while
the other group is attracted by financial advantages.

The number of similar and different neighbors can be easily obtained
from the spin variables of sites neighboring $i$,

\begin{equation}
  N_s(i)-N_d(i)=s_{i} \sum_{\<i,j\>}s_j,
  \label{eq:3}
\end{equation}

\begin{equation}
  N_s(i)+N_d(i)=s_{i}^2\sum_{\<i,j\>}s_j^2,
  \label{eq:4}
\end{equation}
where the sum over $\<i,j\>$ should be understood as a sum over all
$j$ which are neighbors of $i$. Substituing Eq. \eqref{eq:3} and
\eqref{eq:4} into Eq. \eqref{eq:1}, we can rewrite our condition for
the satisfaction of agent $i$, Eq. \eqref{eq:2bis}, as

\begin{equation}
  -s_i\sum_{\<j\>}s_j-
  \(2T-1\)s_i^2\sum_{\<j\>}s_j^2+2Ds_i^2+2Hs_i\leq0,
  \label{eq:5}
\end{equation}
where $j$ runs over his/her eight closest neighbors (\emph{Moore
  neighborhood}). We consider free boundary conditions, so that only
agents at the borders have less than eight neighbors. Besides, we will
only allow moves that either preserve or reduce the dissatisfaction
index for each agent, and their types are fixed. For constant values
of $T$, $D$ and $H$, the energy of the BEG model becomes a Lyapunov
function:
\begin{equation}
\resizebox{7.7cm}{0.30cm}{%
 $\mathcal{H}=-\sum_{\<i,j\>}\(s_is_j+(2T-1)s_i^2s_j^2\)+
  2\sum_i\(Ds_i^2+Hs_i\)$},
  \label{eq:6}
\end{equation}
with an external magnetic field. This Hamiltonian represents a spin-1
Ising model with coupling constant $J=1$, biquadratic exchange of
strength $2T-1$, crystal field of strength $2D$ and a magnetic field
of intensity $2H$, as it can be seen by comparing with
Eq. \eqref{eq:1}.

\subsection{System dynamics}

Agents can enter or leave the $N\times N$ lattice depending on their
dissatisfaction level, Eq. \eqref{eq:2}, and the economic environment
$H$ and $D$ must be explicitly considered.

The system dynamics is similar to the one described in \cite{key-7p},
and can be explained as follows: at each iteration we choose a random
site $i$. If it corresponds to a vacancy we attempt to occupy it with
an agent of a random type. The movement is accepted if the selected
agent presents $I_\dis(i)\leq0$, see Eq. \eqref{eq:2}. On the other
hand, if the selected site is occupied we attempt an internal or an
external change with equal probabilities. For the internal change, a
vacancy is randomly chosen at site $j$ (implying an infinte range
interaction), and the dissatisfaction index for the agent in the
offered place, $I_\dis(j)$, is calculated. The internal change is
accepted only if the dissatisfaction in the new place is preserved or
reduced: $I_\dis(j)\leq I_\dis(i)$. If the external change is chosen,
agent $i$ attempts to leave the system, which will take place if
$I_\dis(i)>0$. Open city systems are controlled by a variety of
parameters: the economic environment variables, $D$ and $H$, and the
tolerance level $T$.


\section{Results and discussion}
\label{sec:results}

We will consider agents of two types in a $100\times100$ square
lattice with free boundary conditions. The neighborhood of any agent
that is not at the border is comprised by its eight closest
neighbors. The initial configuration is random, with red agents, blue
agents and vacancies constituting each $1/3$ of the total system. Let
us stress that the happiness level of the agents depends on the
tolerance value $T$, the economical level of the system $D$, and also
the economic bias for each type of agent, $H$.

The system undergoes the following process: in the first stage, with
fixed values for $T$ and $D$, with $H=0$, the system evolves allowing
both internal and external relocations, until a stationary state is
reached with both agent populations settled inside clusters, as we can
see in Fig. \ref{fig5}. Now, we proceed to increase $H$ gradually, in
order to provide an economically advantageous perspective to one type
of agents over the other. Thus, a gentrification process will
start. As a consequence one of the agent types will tend to leave the
system, while the other will fill these vacancies, giving rise to
avalanches. From this moment on, we will not allow internal
relocations, because our aim is to characterize migratory movements
inside and outside the city.

\subsection{Types of borders}
\label{subsec:typeofborder}

\begin{figure}
\centering
\begin{tabular}{cc}
\includegraphics[width=3.5cm]{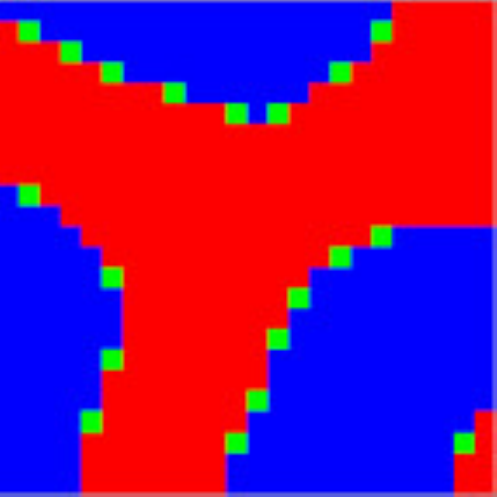} &
\includegraphics[width=3.5cm]{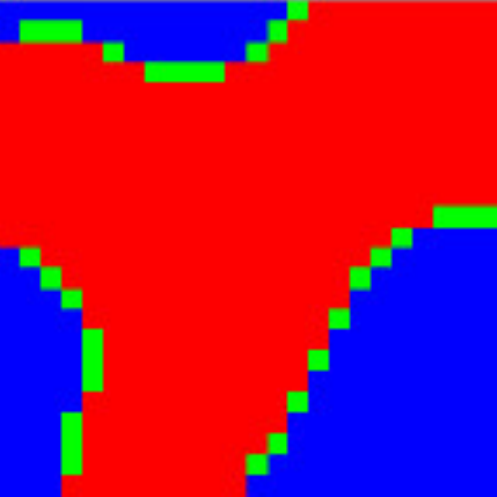} \\
(a) & (b)\\
\includegraphics[width=3.5cm]{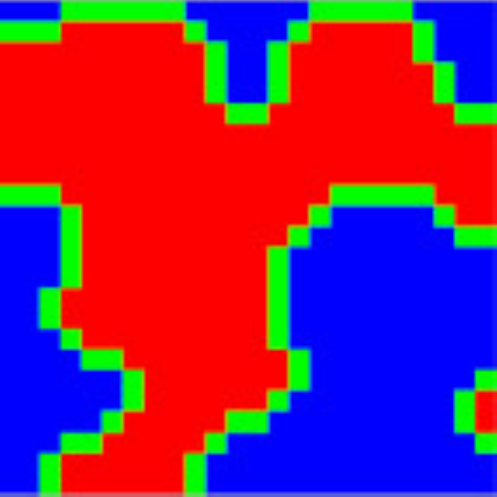} &
\includegraphics[width=3.5cm]{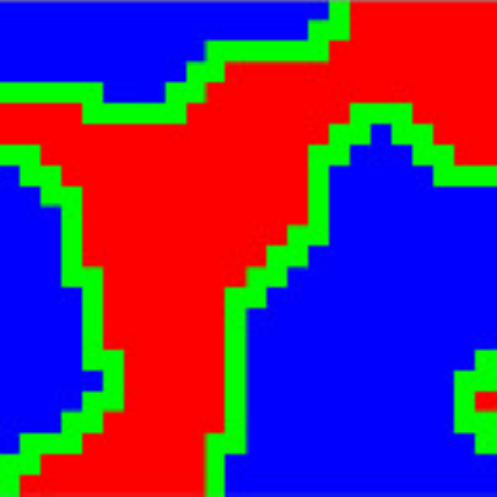} \\
(c) & (d) \\
\end{tabular}
\caption{For $T=1/4$, detail of the border types:
  C (a), 0 (b), I (c) and II (d). Green squares represent vacancies.}
\label{fig1}
\end{figure}

Starting from a random configuration, we set a finite value for $T$
and $D$, with $H=0$. Given enough time, the system reaches an
equilibrium state where the borders have both straight and curved
segments. In this scenario, $N_s$ can be either $4$ or $5$ for agents
at the boundary, depending on whether the agent stands at a straight
or a curved point of the boundary, as we can readily see in
Fig. \ref{fig1}. The dissatisfaction index of each agent disminishes
when the number of similar neighbors increases (see
Eq. \eqref{eq:2}). Thus, any agent on a curved spot will leave the
lattice with higher probability than one at a straight spot. This
difference is the key to the creation of different types of
borders. Notice that, for a given value of $T$, an agent for which
$D>T(N_s+N_d)-N_d$ becomes unsatisfied and is transferred out of the
lattice. Thus, different types of borders can arise, associated with
their value for $N_d$.

\begin{itemize}
  
\item $N_d=4$. If $D<0$ the system is economically very advantageous
  for all agents. Thus, there are almost no vacancies in the system
  and they are located in the corners between clusters. We will call
  this a {\em C-type border}, see Fig. \ref{fig1} (a).

\item $N_d=3$. Vacancies appear at straight segments separating both
  types of agents, in addition to corner sites (Fig. \ref{fig1}
  (b)). This kind of interface location will be termed a {\em 0-type
    border}.

\item $N_d=2$. Straight segments of vacancies are completed, but
  contacts along diagonals between distinct clusters are allowed
  (I-type border \cite{key-7p}), as in Fig. \ref{fig1} (c).

\item $N_d=1$. Diagonal contacts between clusters are not allowed
  any more (II-type border \cite{key-7p}), see Fig. \ref{fig1} (d).
  
\item $N_d=0$. This is the usual situation in economically handicapped
  systems, where $D>0$. When $D$ is above this treshold the system
  expels all agents.
  
\end{itemize}

In this paper we have chosen to determine the frontier between
clusters via threshold values for the number of neighbors of different
types, leading us to define borders of type C and 0, which have not
been reported previously. We have opted for the term C because they
form at the corners between different clusters, see Fig.\ref{fig1}
(a). Departing from the C-type, and before reaching the complete
configuration of a type-I site, we have found an intermediate border
state defined by us as the 0-type, see Fig.\ref{fig1} (b). For type I
and II we have conserved the nomenclature of \cite{key-7p}. As it is
also explained in \cite{key-7p}, these types of borders also appear
for other combinations of parameters $T$ and $D$, so we have used
$T=1/4$ as an illustration.

\subsection{Avalanche processes}
\label{subsec:avalprocop}

At this point, we proceed to increase the economic bias $H$ gradually
by a fixed amount $\Delta H \gtrsim T/2$ every 50 time-steps. This
value is approximately half of the one needed to start any avalanche
process for a given neighborhood configuration. In the economic
interpretation, the system offers less and less financial perspectives
to blue agents, whose economic level and prospects are given by
$\mu_{b}=D+H$. Meanwhile, red agents become more and more economically
favored, $\mu_{r}=D-H$, thus enlarging the economic gap. Both
expressions account for the dissatisfaction of an agent due to their
economic conditions and their perspectives. Moreover, if we define the
happiness associated to being in a specific neighborhood as
$\lambda=T(N_s+N_d)-N_d$, the satisfaction condition,
Eq. \eqref{eq:2}, can be expressed now as

\begin{equation}
  \mu_{r,b} \leq \lambda.
  \label{eq:13}
\end{equation}

The interpretation of Eq. \eqref{eq:13} is straightforward: when the
satisfaction of being into a neighborhood, $\lambda$ compensates the
unhapiness arising from the economic situation, $\mu_{r,b}$, the agent
remains in the system. As we keep increasing $H$, the economic bias
between $\mu_{r}$ and $\mu_{b}$ opens up. The system becomes less
attractive towards blue agents, which increase $\mu_{b}$, and some of
them are forced to leave when Eq. \eqref{eq:13} ceases to
hold. Therefore, there appear some new vacancies in the lattice. At
the same time, the effect is opposite for red agents: $\mu_{r}$
decreases and the lattice is more satisfactory for them, because they
are getting more and more economic advantages. Thus, red agents come
from outside and occupy the vacancies previously created. Now, some of
the blue agents close to these occupied locations cease to verify
Eq. \eqref{eq:13} because $\lambda$ has decreased for them due to the
arrival of red agents. Thus, they may be transferred out of the system
on subsequent time-steps. This gives rise to further new vacancies
that are filled again with red agents and the process goes on in a
self-sustained way. This is what we called a {\em blue avalanche},
because it originates with blue agents leaving the lattice.

Yet, there is another way to generate an avalanche. We depart from an
equilibrium situation with some vacancies, which requires a mid-ranged
economic environment, check Fig. \ref{fig1} (b), (c) and (d). Before
$H$ is strong enough to force blue agents out, red agents may be able
to fill these vacancies up. Blue agents next to these locations may
cease to verify Eq. \eqref{eq:13}, as in the previous situation, and
are forced out of the system. These new vacancies are occupied by red
agents from outside, making further blue agents leave. The process
goes on, giving rise to what we will call a {\em red avalanche}.

The value of $H$ remains constant while these avalanches take
place. After the avalanches are finished, if there are still some blue
agents in the system the value of $H$ is increased further until a new
avalanche takes place.

The similarity with a gentrification process is clear: one of the
types of ageents find economical advantages by moving into this city
zone, while the other is suffering the lack of financial opportunities
and is forced out of the system, leaving vacancies on the borders
between the two clusters. Meanwhile, red agents with better economic
perspectives enter the system and occupy these vacancies. The process
becomes self-sustained because other agents from the less favoured
group are forced to leave the system due to their proximity to members
of the other class. Of course, not all processes in the system give
rise to an avalanche.

Avalanches will be characterized by their size $s$, defined as the
total number of blue agents that have left the lattice as a
consequence of the departure of the first one. We have obtained the
avalanche size histograms, and fitted them to a probability density
function (PDF) of the form

\begin{equation}
  p(s)=Cx^{\alpha}\exp(-x/x_0),
  \label{eq:ps}
\end{equation}
where $C$ is the normalization constant, $\alpha$ accounts for the
scaling exponent for the avalanches and $x_0$ acts as a maximal cutoff
value. For reasons of numerical stability we focus on the
complementary cumulative distribution function (CCDF) \cite{key-16p},
which is fitted by the expression
$C^{*}x^{\alpha+1}\exp-(x/x_{0})$. Data from 100 complete extinctions
of the blue agents in the system have been measured for each $D$
value. For more numerical details, see \ref{app}.

The next sections are devoted to the analysis of avalanche
distributions for three values of $T$: low ($T=1/4$), medium ($T=1/2$)
and high ($T=3/4$). For each value of $T$ we have characterized the
behavior of the system for a wide range of fixed values of $D$,
ranging from the situation where no vacancies are present to the
extinction point in which, due to the high hostility of the medium all
agents leave. We will vary $H$ in order to observe the avalanche
processes for each $D$ value considered.

Let us introduce a convenient notation to describe neighborhoods,
using {\em s} and {\em d} to denote similar and different neighbors,
respectively, e.g. $3s+2d$ means that a certain agent has 3 similar
and 2 dissimilar neighbors.

\subsection{Avalanche distributions for $T=1/4$}
\label{aval14}

We begin our study in a city with a high economic interest, $D=-2.125$
and low tolerance value $T=1/4$.

Under these circumstances the system exhibits a stationary state with
big clusters of red and blue agents and no vacancies, due to the low
value of $D$. Now, we increase $H$ gradually, $\mu_{b}$ becomes
higher, and some blue agents may abandon the lattice. In fact, these
blues agents are the ones which do not verify Eq. \ref{eq:13}, and
have $4s+4d$ neighbors. It must be noted that knowing the value of $T$
and the number of different and similar neighbors, $\lambda$ is
fixed. Meanwhile, the situation for red agents keeps improving:
$\mu_{r}$ becomes lower, so the vacancies created by the blue agents
are filled by red agents coming from outside. This is what we defined
previously as a blue avalanche (Sec. \ref{subsec:avalprocop}).
	
\begin{figure*}
\begin{tabular}{cc}
\includegraphics[width=7.5cm]{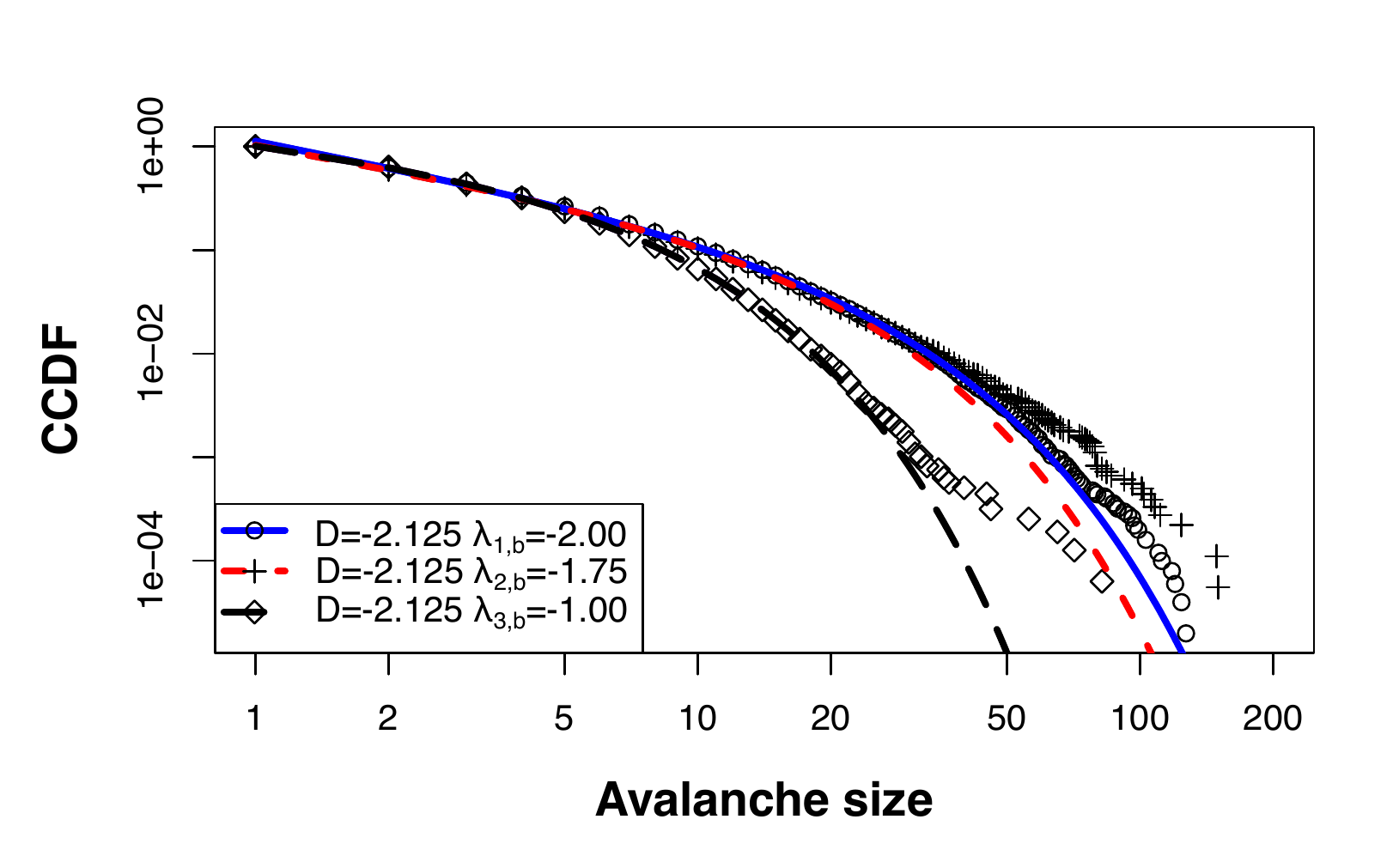} &
\includegraphics[width=7.5cm]{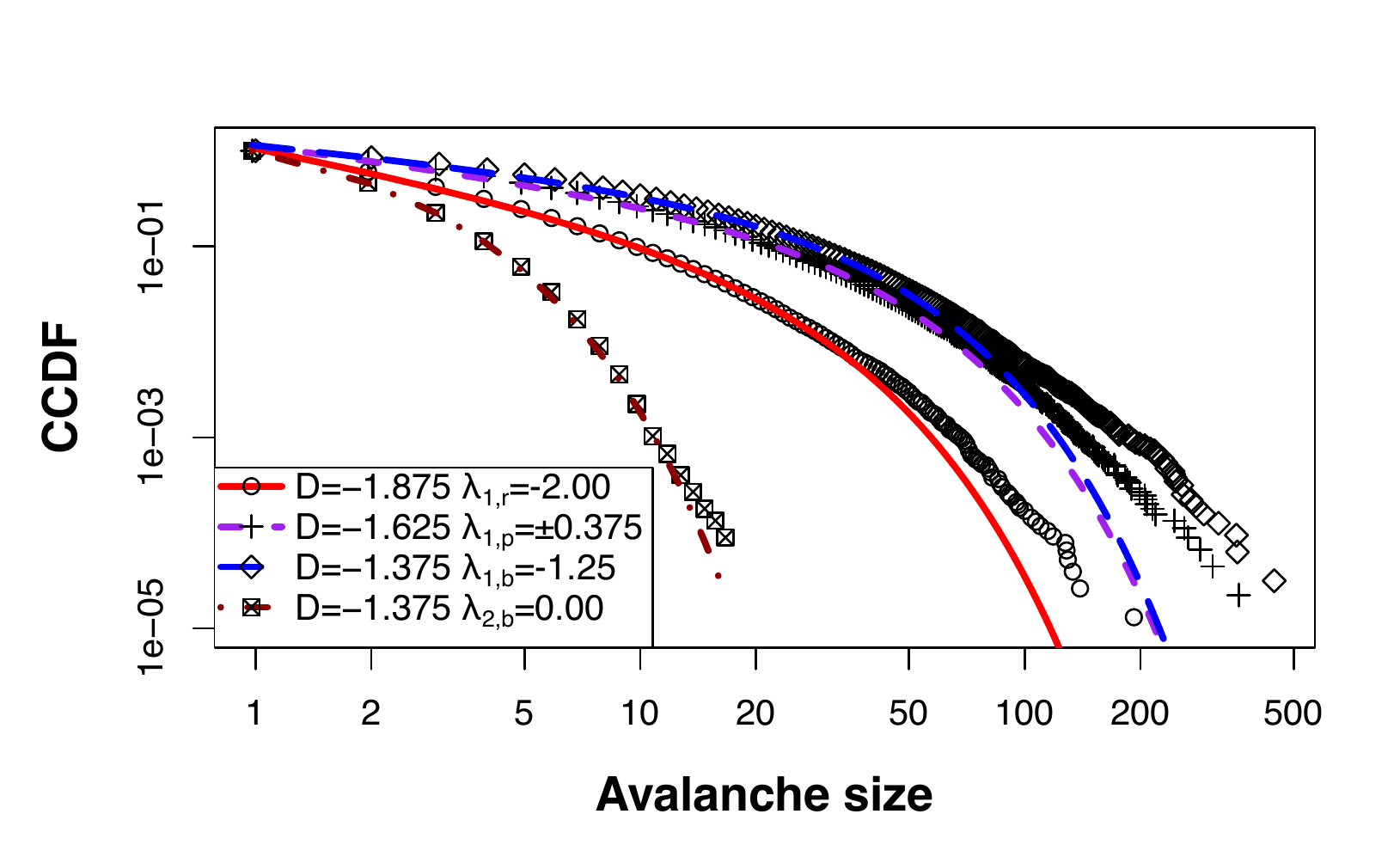}\\
(a) & (b)\\
\end{tabular}
\caption{CCDF of the avalanche distribution sizes for $T=1/4$. For
  each curve the $D$ value is specified. Treshold values are given as
  $\lambda_{n,k}$ where $n$ is the avalanche set index and $k$ its
  kind: $r$ for reds, $b$ for blues and $p$ for purple ones. For
  purple avalanches upper and lower tresholds are expressed as $D\pm
  \lambda_{1,p}$. The fitted power-law cutoff functions are depicted
  with lines.}
\label{fig2}
\end{figure*}

The system is so interesting economically for blue agents that they
will adopt different strategies in order to remain, such as the
formation of special cluster configurations: triangles and rectangles
in contact with the system border. Thus, up to three avalanche
processes can be needed to deplete the system of blue agents, as can
be seen in Fig. \ref{fig2} (a), where $D=-2.125$. To clarify the order
in which each avalanche occurs and its kind (blue or red), we use the
notation $\lambda_{n,k}$, where $n$ denotes the order and $k$ takes
value $r$ for red and $b$ for blue ones. The fitted values for
$\alpha$ and $x_{0}$ in all these cases are shown in Table
\ref{table1}.

For $D=-2.125$ and increasing $H$ gradually, we find that the first
three avalanches start at treshold values $\lambda_{1,b}=-2.0$,
$\lambda_{2,b}=-1.75$ and $\lambda_{3,b}=-1.0$. Extending our
notation, we may say that each avalanche set has a dominant process,
which is characterized by the typical neighborhood of the expelled
blue agents and their $\lambda$ value. The values previously
calcultaed correspond with certain neighborhood types: $4s+4d$,
$2s+3d$ and $5s+3d$, respectively. Of course, no avalanche consists of
a single type of process in practice. We must note that when blue
agents with a higher $\lambda$ start their expulsion (such as
$2s+3d$) those with an inferior treshold (for example $4s+4d$) can be
still be in process of leaving. Thus, it is possible for various
processes to take place simultaneously in the same avalanche. This is
the situation for the first and second avalanche sets, $\lambda_{1,b}$
and $\lambda_{2,b}$ in Fig. \ref{fig2} (a). The process $4s+4d$
dominates both avalanches so their curves are close to each other. As
the avalanche for $\lambda_{3,b}$ is the last one for this $D$ value,
it has a smaller size due the small number of blue agents remaining on
the lattice.

The social meaning is clear: the less favoured agents will try to
stand on an economic advantageous environment despite their economic
gap with the other group. To achieve this goal diverse neighborhood
structures will be created ({\em ghettos}). This is the case scenario
for neighborhoods as Harlem or Clinton Hills \cite*{Harlem}.

Now, let us focus our attention on higher values of $D$: $-1.875$,
$-1.625$ and $-1.375$, as shown in Fig. \ref{fig2} (b). Although the
environment is still economically advantageous, the economic
conditions are not as good as before. Some vacancies appear in the
lattice and the system presents a C type of border (Fig. \ref{fig1}
(a)). As we explained before, different kinds of avalanches are still
possible. The first one takes place for $D=-1.875$. Red agents coming
from outside fill the existing vacancies on the system when
$\mu_{r}\lesssim -2.0$. That points at a $4s+4d$ red avalanche with
$\lambda_{1,r}=-2.00$ (Sec. \ref{subsec:avalprocop}). The next one,
$D=-1.625$, is what we may call a \emph{purple avalanche}: a
simultaneous red avalanche with $\lambda_{1,r}=-2.00$ as in the former
case, and a blue avalanche, $4s+3d$ with $\lambda_{1,b}=-1.25$, due to
the overcoming of both tresholds at the same time ($D\pm H$). We must
note that the $4s+3d$ is the vecindary associated with a 0-type border
(Fig. \ref{fig1}(b)). So, while the red agents coming from outside are
filling the vacancies with $4s+4d$ which correspond to a C-type
border, blue agents with $4s+3d$ are leaving the system creating a
0-type border. Finally, we have blue avalanches for the first and
second avalanche sets with $D=-1.375$, being their associated
neighborhoods $4s+3d$ and $3s+1d$, respectively. The behavior of blue
avalanches is similar to the ones explained for
$D=-2.125$. Interestingly, red avalanches are more frequent than
blue ones, although they present smaller sizes, as we can readily see
in Fig. \ref{fig2} (b) and Table \ref{table1}. Red avalanches
interfere among them, thus reducing the maximum cut-off, $x_0$.

\begin{figure*}
\begin{tabular}{ccc}
\includegraphics[width=5cm]{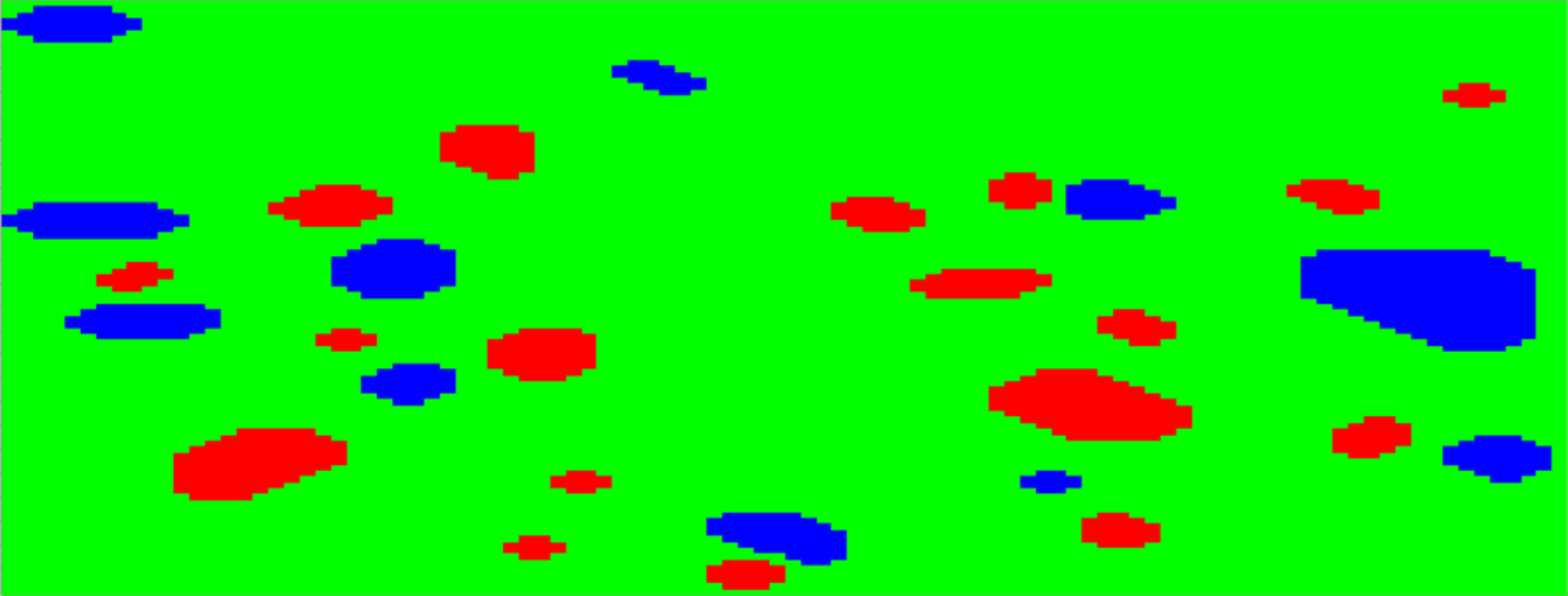} &
\includegraphics[width=5cm]{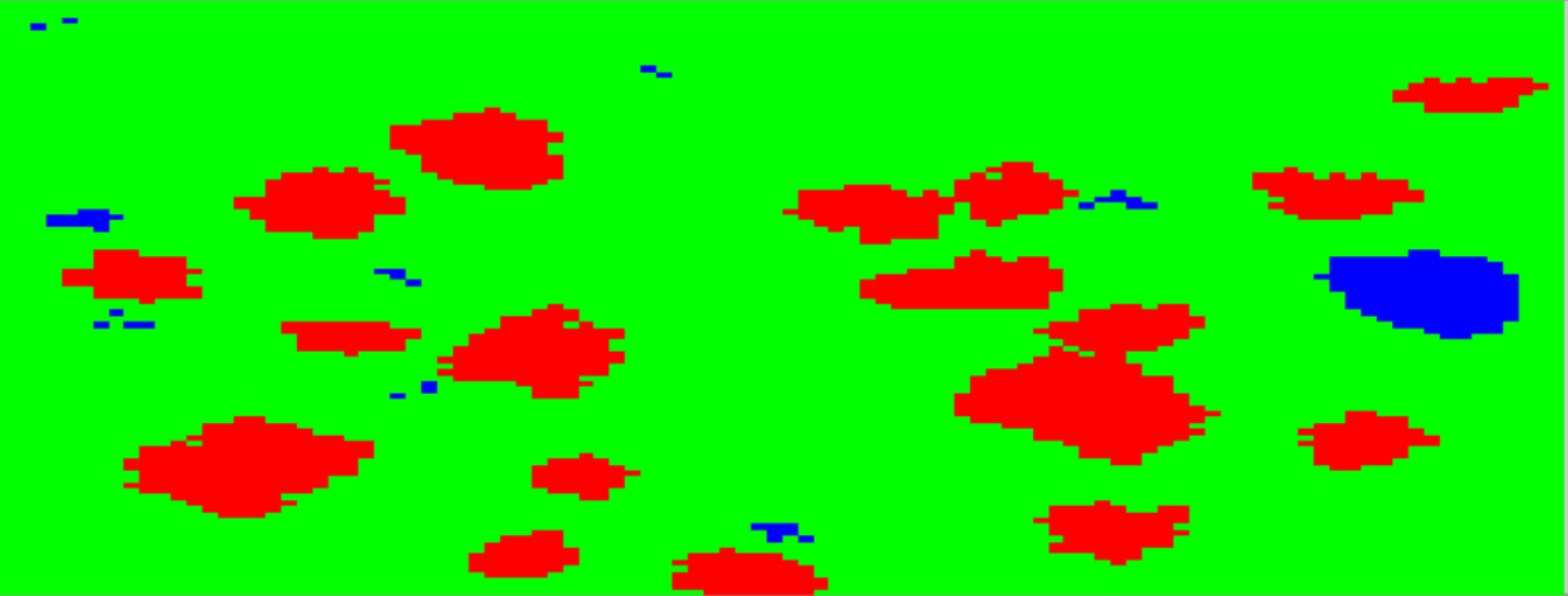} &
\includegraphics[width=5cm]{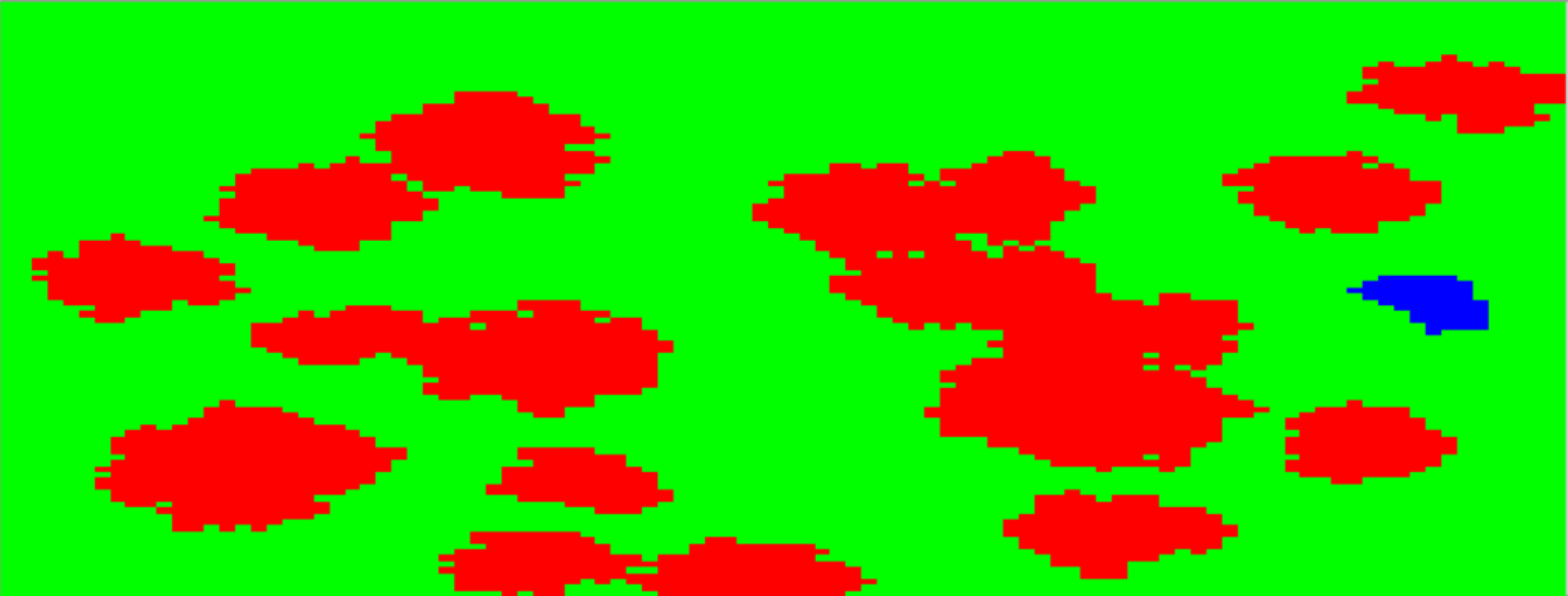} \\
(a) & (b) & (c)\\
\end{tabular}
\caption{From left to right: snapshot of the system evolution for
  T=1/4 and D=0.875. Equilibrium (a), 6 MC steps (b) and 13 MC steps
  (c). Green squares represent vacancies.}
\label{fig3}
\end{figure*}

The last value that we analyze for $T=1/4$ is $D=0.875$. The behaviour
of the blue avalanche suffers an important change because the
condition for a {\em predominant vacancy state} is verified, $D/T>3$
\cite{key-7p}. Combining eq. \eqref{eq:2}, with $H=0$, and Eq. \eqref{eq:2bis} the condition for an agent to remain on the system  
    can be written as $N_{d} \leq (N_{s}-D/T)/(1/T-1)$. As $T$ is lower than unity, the denominator stays positive, so the numerator of the former equation dictates the agent behaviour. Any agent must fulfill the condition $N_{s} \geq D/T$ to remain on the lattice, even if no different agent is in the vecindary. As the system initial
    configuration is random and both kinds of agents and vacancies are
    equally likely, the probability for an agent to have more than
    three similar neighbors is small. Therefore, when $D/T>3$ the system becomes
    very hostile, agents leave massively, and only a small number of
    them remain inside clusters, shortening $x_0$ (Fig \ref{fig3}
    a). The border between clusters is not relevant anymore: red
    clusters grow with a vecindary as $(3s+0d)$ and blue ones become smaller when agents have a neighborhood $(4s+0d)$     (Fig \ref{fig3} (b) and (c)).

Socially, the situation for blue agents might be compared to the
Chicago suburbs where the population could increase their personal
ties via a community network \cite{key-7p}. These ties may prevent a
massive exodus despite the lack of attractiveness of the
environment. However, in our model, the evolution of the system
departs from this situation and develops in two opposite directions:
blue agents are removed from the system due to the hostile environment
and the lack of economic prospects, in contrast to red agents, which
increase their population finding good opportunities. This could be
understood as the unrelated behavior of two communities. One of them
chooses to cooperate and the ensuing growth overcomes economical
hardships. The other one is unable to strengthen their links and leaves
the city.

Finally, we show the values of the fitted parameters in Table
\ref{table1}, being $x_{c}$ the choosen size for the fitting (\ref{app}). We can find $\alpha$ values in the range $[-1.78, -0.98]$
previously reported in references \cite{key-19p,key-20p,key-22p,key-23p}, concerning self-organized criticality in different systems.

\begin{table}[H]
\begin{centering}
\resizebox{7.8cm}{2.6cm}{%
\begin{tabular}{|c|c|c|c|c|c|c|}
\hline 
$B$&$A$  & $D$ & $\alpha$ & $x_0$ & $x_c$ & $C$\\
\hline 
\hline 
N & b &$-2.125$ & $-1.781\pm0.002$ & $16.2\pm0.3$ & $40$ & $1.20\pm0.03$\\
\hline 
C & r & $-1.875$ & $-1.77\pm0.02$ & $15.0\pm0.3$ & $30$ & $1.15\pm0.02$\\
\hline 
C & p & $-1.625$ & $-1.50\pm0.01$ & $24.7\pm0.6$ & $50$ & $1.20\pm0.03$\\
\hline 
C & b &$-1.375$ & $-1.38\pm0.02$ & $23.7\pm0.5$ & $50$ & $1.21\pm0.03$\\
\hline 
0 & r &$-1.125$ & $-1.37\pm0.03$ & $3.30\pm0.04$ & $20$ & $1.42\pm0.04$\\
\hline 
0 & p &$-0.875$ & $-1.716\pm0.008$ & $42.8\pm0.9$ & $50$ & $1.13\pm0.02$\\
\hline 
0 & b &$-0.625$ & $-1.618\pm0.004$ & $140\pm1$ & $240$ & $1.23\pm0.02$\\
\hline 
I & r &$-0.375$ & $-1.35\pm0.02$ & $18.0\pm0.3$ & $40$ & $1.14\pm0.02$\\
\hline 
I & p &$-0.125$ & $-1.463\pm0.004$ & $79.2\pm0.6$ & $150$ & $1.15\pm0.01$\\
\hline 
I & b &$0.125$ & $-1.495\pm0.005$ & $181\pm4$ & $200$ & $1.26\pm0.02$\\
\hline 
II & r &$0.375$ & $-1.136\pm0.007$ & $20.9\pm0.2$ & $50$ & $1.12\pm0.01$\\
\hline 
II & p &$0.625$ & $-1.41\pm0.03$ & $69.9\pm0.4$ & $150$ & $1.081\pm0.008$\\
\hline 
V & rb &$0.875$ & $-1.33\pm0.02$ & $9.6\pm0.3$ & $20$ & $1.15\pm0.02$\\
\hline 
\end{tabular}}
\par\end{centering}
\caption{Avalanche distribution parameters $(T=1/4)$ for $D$ values
  from (Section \ref{aval14}). $B$ is the type of border, with 'V'
  meaning vacancy dominated regime. $A$ is the avalanche type, ('b' is
  blue, 'p' is purple and 'r' is red), $\alpha$ is the power law
  exponent, $x_{0}$ accounts for cutoff value and $x_{c}$ is the
  chosen size for the fitting (see \ref {app}).}
\label{table1}
\end{table}

\subsection{Avalanche distributions for $T=1/2$ and $T=3/4$}
\label{aval1234}

Most of the processes that arise for these values of $T$ are
avalanches with similar parameter values and behaviours to those
explained in the previous section (Sec. \ref{aval14}).

Nevertheless, as this $T$ fixed value is larger than before, the
dissatisfaction index disminishes and Eq. \ref{eq:13} is verified for
higher values of $D$. Thus, despite being in a less economically
interesting system, agents populate the lattice. As a consequence the
range of values of $D$ for which the system is vacancy dominated
increases.

\begin{figure}
\includegraphics[width=8cm]{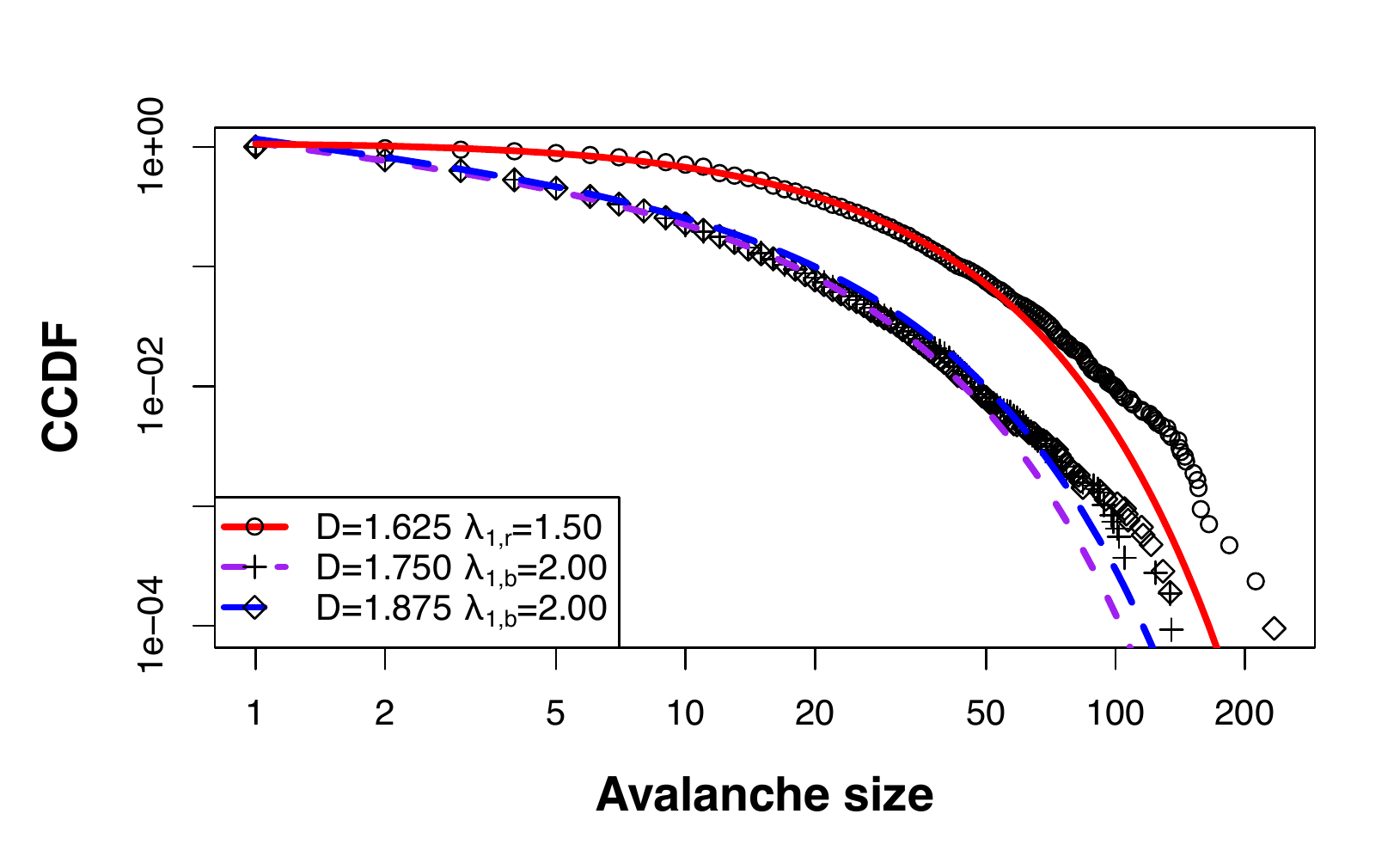}
\caption{CCDF of the avalanche distributions for $T=1/2$ in the
  predominant vacancy state, represented by symbols. The fitted
  power-law cutoff functions are depicted with lines.}
\label{fig4}
\end{figure}

The two last avalanches for $T=1/2$, $D=1.750$ (short dash) and
$D=1.875$ (long dash) are in the predominant vacancy state (Fig. \ref{fig4}). They are
also really close to each other, suggesting that the same mechanism
takes place and it is related to $4s+0d$ neighborhoods with
$\lambda_{1,b}=2.00$. This process implies that the
blue agents will be expelled when they are surrounded by vacancies.

The system evolution for $D=1.750$ is apparently similar to the one in
Fig. \ref{fig3}. Nonetheless, the environment has become so hard that
red agents are not able to expand as the blue agents progressively
leave the city, converting a purple avalanche into a blue one. From a
social perspective it can be explained as people leaving the city
looking for a better financial situation somewhere else, abandoning
the neighborhood into obliteration.

\begin{figure*}
\begin{center}
\begin{tabular}{ccc}
\includegraphics[width=5cm]{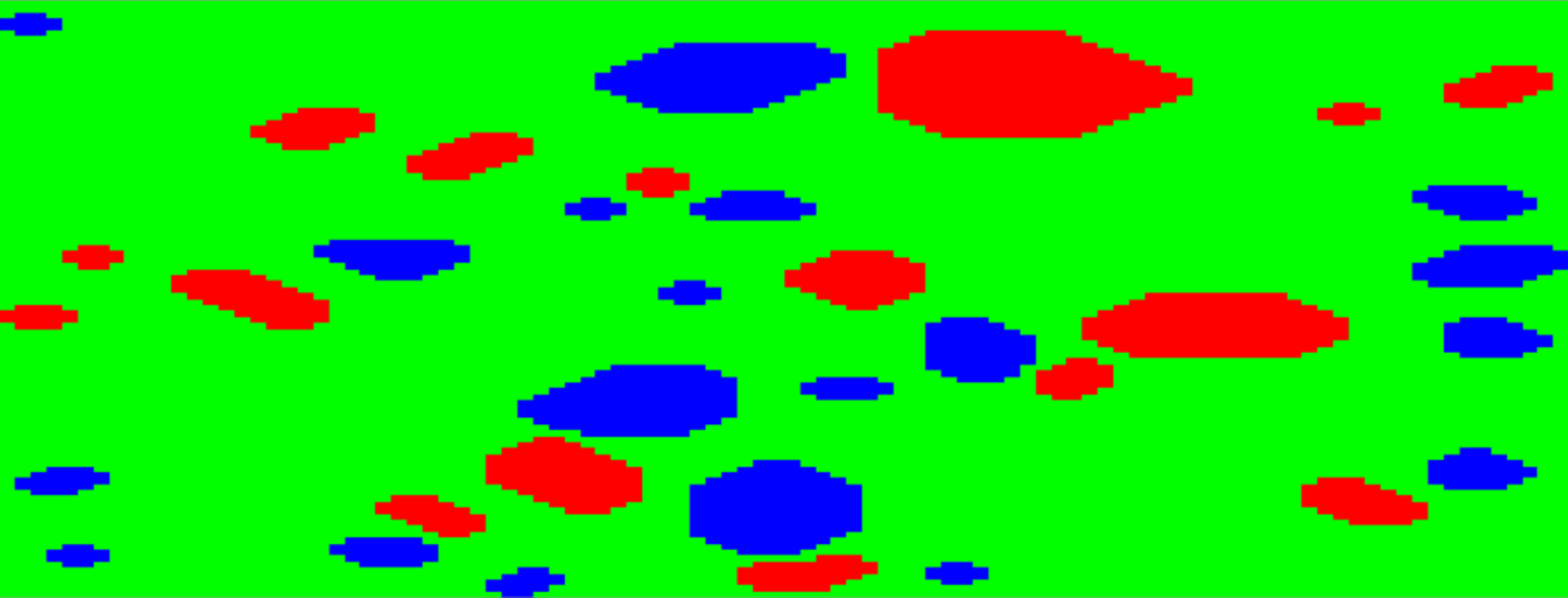} &
\includegraphics[width=5cm]{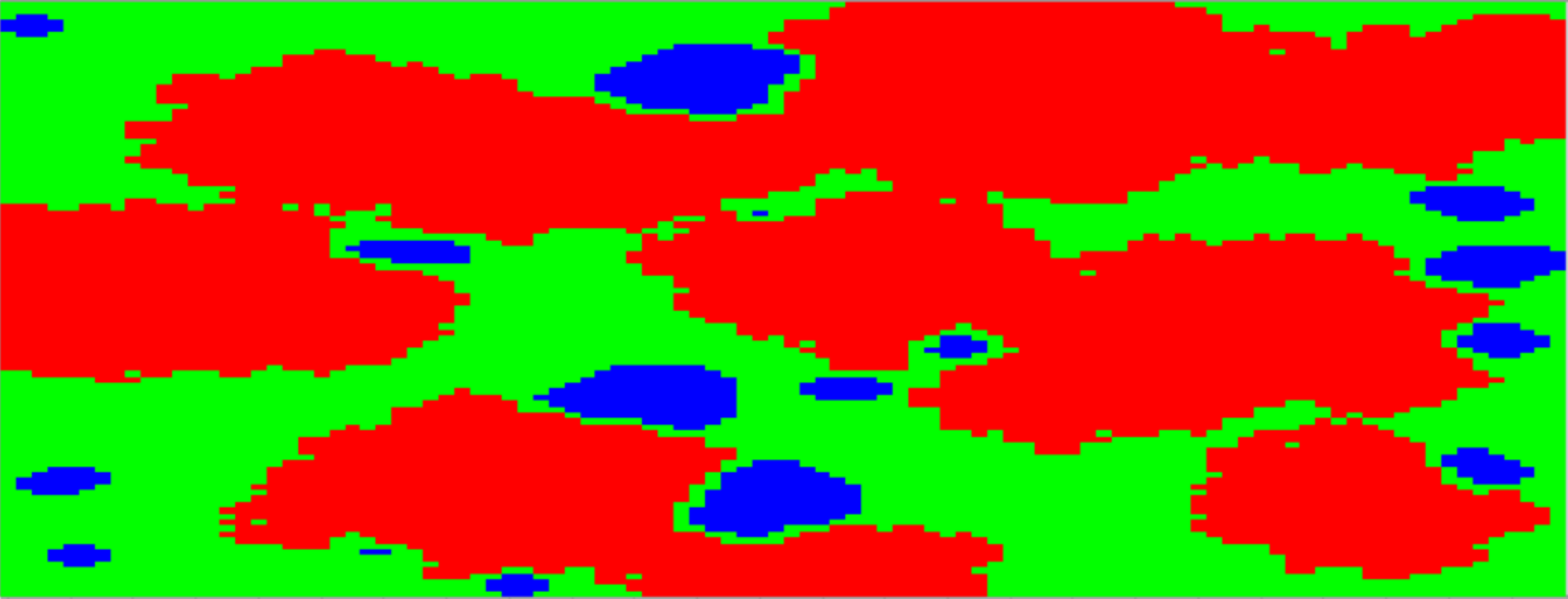} &
\includegraphics[width=5cm]{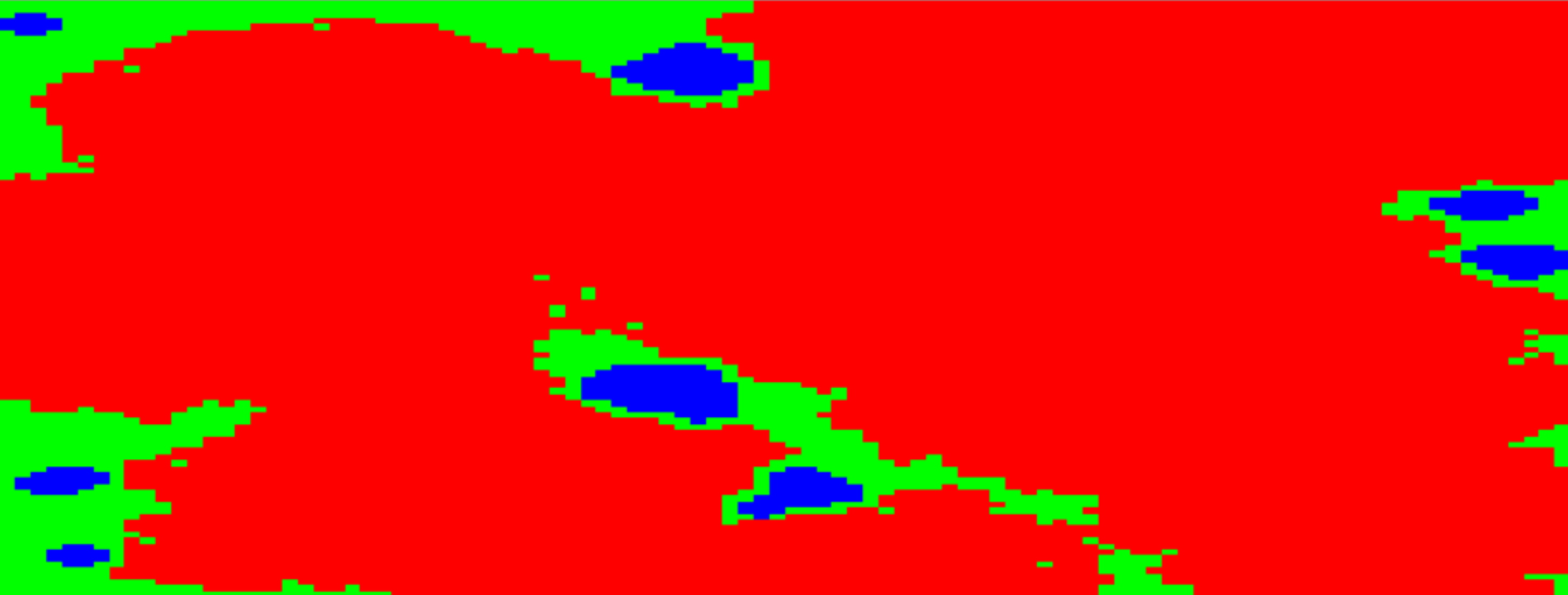} \\
(a) & (b) & (c)\\
\end{tabular}
\par\end{center}%
\caption{From left to right: snapshot of the system evolution with
  $T=1/2$ and $D=1.625$. Equilibrium (a), 30 MC steps (b) and 50 MC
  steps (c).}
\label{fig5}
\end{figure*}

One of the most interesting processes in our work takes place for
$D=1.625$ (continuous), and is also found in the predominant vacancy
state (Fig. \ref{fig4}). It is convenient to depict its evolution in
Fig. \ref{fig5}. As it is shown, the first procedure that takes place
is the red cluster expansion (Fig. \ref{fig5} b)) for the
neighborhood $3s+0d$ which takes place when $\mu_{r} \lesssim
1.5$. This mechanism proceeds until a blue cluster is found. As the
red cluster grows, the population increases. When both kinds of agents
are close, the $4s+1d$ red avalanche removes all blue agents. For this
$D$ value, $\alpha=-0.97\pm0.01$, $x_{0}=17.4\pm0.3$.

This two stage process may be interpreted as the expansion of some
kind of neighborhoods that improve their economical perspectives and are able to increase their size, wrapping around other devaluated
zones. After that, a gentrification process starts. This process is
reminiscent of the transformation of small towns or villages close to
growing commuting zones \cite{key-24p}.

Finally, for $T=3/4$, multiple $D$ values are found in the vacancy
dominated regime, but all the situations are related to those
previously explained. Type of borders 0, I and II are found to be in
this situation.

When the system is in the predominant vacancy state, there are only
two processes that guarantee the complete removal of blue agents. The
first one is associated with $3s+0d$ neighborhoods and consist of the
red cluster expansion, which changes the environment. After that,
other complementary processes take place (as in Fig. \ref{fig5}). The
other one is related to $4s+0d$ neighborhoods, which implies the total
expulsion of blue agents surrounded by vacancies (Fig. \ref{fig3}). As
a consequence, fitted parameters are close to the ones previously
calculated for other values of $T$ in the same situations.


\section{Conclusion}
\label{sec:conclusion}

The generalization of the open city model provides a new
framework for the study and understanding of a broad class of
urban processes, i.e. gentrification. Besides analyzing its results from a social and economical perspective, the model is also linked with the physics of the BEG model under the influence of an external magnetic field. 

As agents can leave or enter the city, not only tolerance but two
economic terms are taken into consideration: $D$ is associated with
the mean economic city level, and $H$ stands for the economic
attractivity gap between both types of agents. The dynamical rules are
the following: once the system has evolved into an equilibrium state,
fixing $T$ and $D$, $H$ progressively increases. In the economical
sense, there are worse financial perspectives for blue agents,
meanwhile, red agents are finding better opportunities, and an
economic gap is created. Some blue agents, generally the ones that are
closer to the red agents, begin to leave the city. These new vacancies
are occupied by red agents coming from outside, so the process goes on
in a self-sustained way, and resembles gentrification.

A power law with exponential cutoff expression has been used to
describe the avalanche size histograms. While the cutoff length
depends on the system size, the power law exponents are in the range
$[-1.781,-0.97]$, values that can be found in the literature for
diverse avalanche processes.

In our modified Schelling model gentrification processes could also
help understand the formation of ghettos, as special configuration of
the less favoured class in order to remain in the cities. Moreover,
our results highlight the importance of tolerance and ties for people
to be satisfied despite harsh conditions. On one hand, collaboration
inside of a neighborhood implies an improving in economic and trading
exchanges, making growth possible. On the other hand the lack of economic stimulus in a disfavorable environment results in the neighborhood progressive degradation.

Further analysis should focus on variants in which agents consider
different economic zones. For example, the city centre is usually more
valuable than the outskirts. Furthermore, it could be also interesting
to study the distribution of the inter-event times between the
avalanches shown in this work. Finally, we consider that the presented
framework could also allow us to study a recent urban phenomena: the
relocation of workers with telecommuting possibilities outside the
cities due to COVID19 pandemic.

\section*{Acknowledgments}
  We acknowledge financial support from the Spanish Government through
  grants PGC2018-094763-B-I00 and PID2019-105182GB-I00. We would also
  like to thank the referee for giving such constructive comments
  which helped to improving different aspects from this work.

\bibliography{openbib2}

\begin{appendix}

\section{Fitting the avalanche histograms}
\label{app}

Some numerical difficulties associated with direct estimation of
$\alpha$ and $x_{0}$ from the probability density function (PDF) of
the histograms are known to arise \cite{key-16p}. In order to address
them, we have resorted to the use of the complementary cumulative
distribution function (CCDF). Our analytical form is given by
Eq. \eqref{eq:16}.

\begin{equation}
\begin{split}
P(s) &=Pr(X>s)\\
& = \int_{s}^{\infty}Cx^{\alpha}\exp(-x/x_{0})dx  \\
& = F(\alpha,x_{0})+G(\alpha,x_{0})s^{\alpha+1}\exp-(s/x_{0}) \\
& \cdot \left[1+\frac{1}{2+\alpha}\frac{s}{x_{0}}+...\right].
\end{split}
\label{eq:16}
\end{equation}

The terms inside the brackets are an expansion of
$\mathbf{M}(1,2+\alpha,s/x_{0})$, where $\mathbf{M}$ is the confluent
hypergeometric function. Once $\alpha$ and $x_{0}$ have been
estimated, both $F(\alpha,x_{0})$ and $G(\alpha,x_{0})$ take fixed
values. In fact, $F(\alpha,x_{0})$ does not introduce significative
changes in the fit, and it can be neglected. Under these
circumstances, and retaining only the leading term of the series, we
have

\begin{equation}
  P(s)\approx G(\alpha,x_{0})x^{\alpha+1}\exp(-x/x_0).
\end{equation}

Consequently, we can infer the exponent $\alpha$ and the length $x_0$
of an avalanche from the CCDF of the avalanche histogram.

The experimental CCDF data series in our avalanches have been
calculated from 100 complete extinctions of the blue agents. After
that, data are plotted with a constant bin size by means of the {\tt
poweRlaw} R package \cite{key-17p}. Then an expression of the type
$C^{*}x^{\alpha+1}\exp-(x/x_{0})$, where $C^{*}\approx
G(\alpha,x_{0})$ is fitted by the nonlinear least square method from
the {\tt stats} subroutines \cite{key-18p}. Although the curves have
been depicted for a wide avalanche size range, they have been fitted
inside the interval $[1,x_{c})$, being $x_{c}$ a choosen value to
uphold precision. Deviations between data and fit can appear for
$s\gg x_{0}$, due to the series expansion approximation and the own
nature of the tail distribution, but they are not relevant. As a
practical rule, if $x_{0}<60$ then $x_{c}\leq2x_{0}$ (see Table
\ref{table1}). Choosing $x_{c}$ in this way, the precision of the
fit improves for a wide range of avalanche size values.
\end{appendix}

\end{document}